\documentclass[12pt]{article}
\usepackage{latexsym,axodraw}

\def\ds{\displaystyle}

\begin{document}
\begin{titlepage}
\begin{flushright} KUL-TF-96/1\\ hep-th/9601095
\end{flushright}
\vfill
\begin{center}
{\large\bf Higher Covariant Derivative Pauli-Villars Regularization
for\\[4pt]
Gauge Theories in the Batalin-Vilkovisky Formalism.}\\[10mm]
{\sc P. Claus$^1$}\\[1.5cm]
{\em Instituut voor Theoretische Fysica,
\\Katholieke Universiteit Leuven,
\\Celestijnenlaan 200D,
\\B-3001 Leuven, Belgium}
\end{center}
\vfill
\begin{center}
{\bf Abstract}
\end{center}

\begin{quote}
{\small
The combined method of Higher Covariant Derivatives and Pauli-Villars
regularization to regularize pure Yang-Mills theories is formulated in the
framework of Batalin and Vilkovisky.  However, BRS invariance is broken by
this method and suitable counterterms should be added to restore it.  The
1-loop counterterm is presented.  Contrary to the scheme of Slavnov,
this method is regularizing and leads to consistent renormalization group
functions, which are the same as those found by other regularization
schemes.}

\vskip5mm \hrule height.1pt width 5.cm \vskip 1.mm
{\small\small
\noindent $^1$Wetenschappelijk Medewerker, NFWO, Belgium;
E-mail: {\tt Piet.Claus@fys.kuleuven.ac.be}}\\ \normalsize
\end{quote}
\begin{flushleft}
Januari 1996
\end{flushleft}
\end{titlepage}

\begin{section}{Introduction}
In order to regularize gauge theories, viz.  Yang-Mills theories, the most
successful and practical regularization method has become dimensional
regularization \cite{dimreg}.  Though for theories, whose properties
explicitly depend on the dimension of spacetime ---like e.g.  chiral gauge
theories or supersymmetric gauge theories--- dimensional regularization
requires careful treatment to say the least.  Moreover one can even not
define the path integral in dimensional regularization and there is no
clear prescription to treat anomalies.  Therefore we have to look for
regularization schemes that explicitly stay at the physical space-time
dimension.

The method of higher covariant derivatives (HCD) \cite{HCDreg} seems well
established.  The advantage of this method is that gauge invariance (and
BRS invariance) can be conserved explicitly.  However the regularization is
incomplete in the sense that it doesn't regularize the one loop theory.  A
second regulator is therefore needed to take care of the one loop
divergencies.  The choice of this regulator is not straightforward.  One
could use dimensional regularization, but this conflicts somehow with the
original motivation to avoid this regularization scheme, as this procedure
can not be extended to gauge theories whose gauge invariance is linked to
the dimensionality of space-time.  Nevertheless it leads to consistent
results \cite{ruizdim}.  Slavnov proposed as second regulator a combination
of Pauli-Villars(PV) determinants \cite{slav}.  This method was believed to
solve the problem as it was thought to regularize the theory staying at the
physical value of the space-time dimension.  However in \cite{Ruiz} it was
pointed out that this method is not regularizing, which is a reason to
discard it as a viable scheme.  One needs a third regulator but the theory
leads to inconsistent results and in the presence of matter even breaks
unitarity\cite{Ruiz, ruizuni}.  It seems that there exists no BRS invariant
HCD~PV~regularization scheme.

However, the renormalizability of a non abelian gauge theory is not tied to
the existence of a regularization scheme that formally preserves BRS
invariance, but to the fact that the BRS symmetry is non
anomalous\cite{pig}.  So one can use a PV~scheme that breaks BRS
invariance.  The calculation of the one loop effective action consists of
several steps.  After the introduction of the PV~fields, one can calculate
the breaking of BRS invariance.  The result for pure Yang-Mills theories
was partially obtained in \cite{JacBRST}.  Then one adds a counterterm to
the action to preserve BRS invariance at the quantum level.  Calculating
with this new action one has a one loop BRS invariant regularized action
and one can proceed like in any BRS invariant regularization scheme.  In
fact the inclusion of the counterterms can be seen as part of the
regularization scheme.

The purpose of this paper is to present this method explicitly for pure
Yang-Mills theories at one loop level, using the PV-method developped in
\cite{PVreg}, combined with HCD.  We show that the inconsistencies of
\cite{Ruiz} are absent.  Also at no point we need to introduce a
preregulator.  Working in the Batalin-Vilkovisky(BV) formalism
\cite{BV,PVreg} to treat gauge theories implies a preferred choice for the
$Z$-factors in multiplicative renormalization, different from the ones used
in standard textbooks \cite{Itz}.

The paper is organized as follows.  In section 2 the HCD regularized action
is derived in the BV~formalism and PV~fields are introduced, according to
\cite{PVreg}.  Finally I comment on multiplicative renormalization in the
BV~scheme.  Section 3 is devoted to the calculation of the one loop
effective action and Renormalization Group coefficients using two
regularization schemes.  First the PV~regularization without HCD is used
and the $Z$-factors are presented in this theory illustrating the problem
outlined above.  This factors are used to calculate the renormalization
group coefficients.  Secondly I present the results for the PV~scheme
combined with HCD.  The conclusions are in section 4.  The Feynman rules
are collected in appendix A.  Appendix B shows a sample calculation using
PV~regularization and explicitly staying at 4 dimensions during the whole
calculation.
\end{section}

\begin{section}{The regularized action \label{s:regac}}
In this section the regularized action is derived.  First the higher
covariant derivatives are introduced in the action and the gauge fixed
action is obtained.  PV~fields are introduced to regularize the one loop
theory.  Finally I comment on multiplicative renormalization.

\begin{subsection}{The Higher Covariant Derivative Regularization}

We consider pure Yang-Mills theorie in 4 Euclidean dimensions and add a
Higher Covariant Derivative (HCD) term:
\begin{equation}
S_{YM}=-\frac{1}{4} F_{\mu \nu}^a(1 - \frac{D^2}{\Lambda^2})^2 F_a^{\mu
\nu}.
\label{Sclass}
\end{equation}

In this action $F_{\mu \nu}^a = \partial_\mu A^a_\nu -\partial_\nu A^a_\mu
+ g f^a_{b c} A_\mu^b A_\nu^c $ is the field strength, $A_\mu ^a$ the gauge
field, $g$ the coupling constant and $f^a_{b c}$ the structure constants of
the gauge algebra.  The covariant derivative is $D^a_{\mu\, b}=
\delta^a_{\, b} \partial_\mu + f^a_{cb} A_\mu ^c$.  In (\ref{Sclass}) a
trace over the algebra indices and an integral over Euclidean space-time is
understood.  I assume that the gauge group is a compact, simple Lie group.
Therefore the structure constants can be taken to be totaly antisymmetric
in the three indices and are normalised so that $f^a_{cd} f^b_{cd} = c_v
\delta^{ab}$ where $c_v$ is the eigenvalue of the quadratic Casimir
operator in the adjoint representation.  For $SU(N)$, $c_v = N$.

In order to construct the quantum theory one needs to fix the gauge and
gauge invariance is turned into BRS invariance \cite{brs}.  In the
following we will use the Batalin-Vilkovisky formalism (BV) \cite{BV,PVreg}
to construct the gauge fixed action.  The field content of the quantum
theory is $\Phi^A = \{ A^a_\mu, c^a, b^a \}$, where $c^a$ will be the
ghosts and $b^a$ the antighosts.  Now antifields are introduced ---which
act as sources for BRS-transformations--- and the extended action is
\begin{equation}
S_{ext}=- \frac{1}{4} F_{\mu \nu}^a (1 - \frac{D^2}{\Lambda ^2})^2 F^{\mu
\nu}_a + A_{a\,\mu} ^* D^\mu c^a + \frac{1}{2} c^{*a} f^a_{bc} c^b c^c -
\frac{1}{2 \alpha } b^{*\, a}b_a^*.
\label{Sclassext}
\end{equation}
This action satisfies the master equation $(S_{ext},S_{ext})=0$, which
expresses the BRS invariance.  Notice the introduction of a non minimal
term $-\frac{1}{2 \alpha} b^{*\, a}b_a^*$ to fix the gauge.  This can be
done by shifting the antifields $A^*_\mu$ and $b^*$ and is in fact a
canonical transformation on the variables $\{\Phi^A, \Phi^*_A\}$ w.r.t.
the antibracket.  Then the gauge fixed action will automatically satisfy
the master equation.\\
As generating function for the gauge fixing canonical transformation we
take $F(\Phi^A,\Phi'^*_A)=\Phi^A\,\Phi'^*_A + b
(1-\frac{\partial^2}{\Lambda^2})\partial_\mu A^\mu $.  This means that the
antifields are redefined as
\begin{equation}
\Phi_A^* = \Phi'^*_A + \frac{\stackrel{\rightarrow}{\partial}}{\partial
\Phi^A} F(\Phi,\Phi'^*).
\end{equation}
This leads to the following extended gauge fixed action
\begin{eqnarray}
S_{gf}(\Phi,\Phi^*)&=&-\frac{1}{4} F_{\mu \nu}^a
(1-\frac{D^2}{\Lambda^2})^2 F^{\mu \nu}_a - \frac{1}{2 \alpha
}(\partial_\mu A^\mu_a )(1-\frac{\partial ^2}{\Lambda^2})(\partial_\nu
A^{a\, \nu}) + b_a \partial_\mu (1-\frac{\partial ^2}{\Lambda^2}) D^\mu c^a
+ \nonumber\\
&&A_{a\, \mu} ^* D^\mu c^a + \frac{1}{2} c^{*a} f^a_{bc} c^b
c^c - \frac{1}{\alpha}b^*_a(1-\frac{\partial ^2}{\Lambda^2})\partial_\mu
A^{a\,\mu} - \frac{1}{2 \alpha } b^*_a b^{*a}.
\label{Sgf}
\end{eqnarray}
This action satisfies the master equation, i.e.  $(S_{gf},S_{gf})=0$ which
implies BRS invariance of $S_{gf}(\Phi,\Phi^*)$ under:
\begin{eqnarray}
\delta_{BRS} A^{a \mu}&=&\left.(A^{a \mu},S)\right
|_{\Phi^*=0}=(D^\mu)^a_{\,b} c^b,\\
\delta_{BRS} b^a&=&\left.(b^a,S)\right
|_{\Phi^*=0}=-\frac{1}{\alpha}(1-\frac{\partial^2}{\Lambda^2}) \partial_\mu
A^{a \mu },\\
\delta_{BRS} c^a&=&\left.(c^a,S)\right |_{\Phi^*=0}=\frac{1}{2} f^a_{bc}
c^b c^c.
\end{eqnarray}
A number of comments are now in order.\\

\noindent
{\em Comment 1} : This theory is finite for all higher loop diagrams,
except for the one loop (sub)diagrams.  If one denotes the superficial
degree of divergence (SDD) for a 1PI diagram $G$ as $\omega (G)$ then one
has for this theory in 4 dimensions that
\begin{equation}
\omega (G)=4 - 4(L-1) - E_A - {\textstyle\frac{7}{2}} E_{gh}
\end{equation}
Now only the one-loop (sub)diagrams with 2,3,4 external gauge fields
$(E_A)$ and 0 external ghosts $(E_{gh})$ are divergent.  So in order to
regularize the one-loop theory one needs another regularization.
Dimensional regularization could be used and everything works out well
\cite{ruizdim}.  However then the introduction of the HCD becomes
superfluous and moreover one of the reasons to look for a HCD
regularization scheme was to avoid the potential problems with dimensional
regularization.\\

\noindent
{\em Comment 2} : The choice of this HCD term leads to propagators which
factorize (see appendix A) and are very convenient in Feynman diagram
calculations because the usual manipulations can be applied.  I also used a
gauge fixing term of the form
\begin{equation}
-\frac{1}{2\alpha}\partial_\mu A_a^\mu(1-\frac{\partial^2}{\Lambda^2})^2
\partial_\nu A^{a\, \nu},
\end{equation}
so that the gluon propagators are behaving like $1/p^6$ (i.e.  the
regularization is not destroyed by the gauge-fixing) and have a factorizing
denominator (see appendix A).  Another advantage of the factorization of
the denominators is that one doesn`t need dimensional regularization
techniques to calculate diagrams.  In ref.  \cite{Ruiz} dimensional
regularization techniques were used not only because the scheme of Slavnov
was not fully regularizing, but also to calculate the diagrams that were
regularized.\\

\noindent
{\em Comment 3} : This regularized theory is manifestly local as
it is obtained from a local functional (\ref{Sclass}) through local
canonical transformations.  In this it differs from the action of Faddeev
and Slavnov \cite{slav} who have a manifestly nonlocal action in the
auxiliary field.  Because we wanted to keep the action manifestly local
there are also higher derivatives in the ghost action and the
BRS-transformation of the antighost.

\end{subsection}

\begin{subsection}{One loop Pauli-Villars regularization \label{ss:PVreg}}
In order to regularize the one loop diagrams I will adopt
PV~regularization.  For the introduction of PV-fields I follow the general
approach of \cite{PVreg}.  Denote first $S_{AB}\equiv
\frac{\stackrel{\rightarrow}{\partial}}{\partial \Phi^A } S
\frac{\stackrel{\leftarrow}{\partial}}{\partial\Phi^B }$.  The PV~action is
given by
\begin{equation}
S_{PV}(\Phi,\Phi^*,\underline \Phi_i)=\frac{1}{2} \sum_{i=1}^{N}{\underline
\Phi}^A_i S_{AB} {\underline \Phi}^B_i - \frac{1}{2} M^2_i {\underline
\Phi}^A_i T_{AB} {\underline \Phi}^B_i, \label{Spv}
\end{equation}
with $\underline \Phi_i$ the PV~copies, $T$ an invertible matrix and $N$
the number of PV~copies needed.  For more information about the statistics
of the PV~fields and the number of copies needed see \cite{PVreg}.  The
BRS-transformation of the PV~fields $\{\underline \Phi^A_i \}$ are defined
as $\delta_{BRS} {\underline \Phi}^A_i \equiv K^A_{\, B} {\underline
\Phi}^B_i$, where $K^A_{\, B}=
\frac{\stackrel{\rightarrow}{\partial}}{\partial \Phi^*_A} S
\frac{\stackrel{\leftarrow}{\partial}}{\partial \Phi^B}$.  This yields
\begin{eqnarray}
\delta_{BRS} {\underline A}^\mu _i&=&\partial_\mu {\underline c}^a_i +
f^a_{bc} A^{b \mu} {\underline c}^c_i + f^a_{bc} {\underline A}^{b \mu}_i
c^c,\label{delpva}\\
\delta _{BRS}{\underline b}^a_i&=&-\frac{1}{\alpha}\partial_\mu {\underline
A}^{a \mu}_i,\\
\delta_{BRS}{\underline c}^a_i&=&f^a_{bc} c^b {\underline
c}^c_i.\label{delpvc}
\end{eqnarray}
For one loop calculations it is very convenient to adopt a
formal\footnote{We can of course introduce PV-fields with a well defined
statistics.  A boson with mass $M$ leads to $c_i=1$ and a fermion pair with
mass $2M$ leads to $c_i=-2$.} integration rule,
\begin{equation}
\int {\cal D} {\underline \Phi}_i \ e^{- \frac{1}{2} {\underline \Phi}^A_i
M_{AB} {\underline \Phi}^B_i} = sDet(M_{AB})^{-\frac{1}{2} c_i},
\end{equation}
where the $c_i$ have to satisfy certain relations in order
to regularize.  The fully regularized path integral is now given by
\begin{equation}
Z_{reg}[J_A,\Phi^*_A] = \int {\cal D} \Phi^A \prod_{i=1}^N {\cal D}
{\underline \Phi}^A_i\, \exp -{\textstyle \frac{1}{\hbar}} \{ S_{gf} +
S_{PV} + J^A \Phi_A\}
\label{Zreg}
\end{equation}
There are some important comments on this PV~action.\\

\noindent
{\em Comment 1} : In this PV~action all the original vertices are copied
and coupled quadratically to the PV~fields, i.e.  every PV~vertex contains
two PV~fields.  This action, apart from the mass-term, is invariant under
the BRS-transformations (\ref{delpva} -- \ref{delpvc}).  Although it is not
necessary to introduce the vertices $b^a (1-\frac{\partial^2}{\Lambda^2})
f^a_{bc} {\underline A}^{b \mu}_i {\underline c}^c_i$ and ${\underline
b}^a_i f^a_{bc}(1-\frac{\partial^2}{\Lambda^2}) {\underline A}^{b \mu}_i
c^c$ for regularization, they are needed for BRS invariance under the
defined transformations for the PV~fields.  With these transformations it
is obvious that the measure in the path integral is invariant if
$\sum_{i=1}^{N} c_i=-1$.  The BRS-variation of the total measure in the
path integral is
\begin{equation}
\delta_{BRS}\left\{{\cal D} \Phi^A
\prod_{i=1}^N {\cal D} {\underline \Phi}^A_i \right\}= \left(1 +
\sum_{i=1}^N c_i \right)K^A_{\, A} = 0.
\end{equation}
In order to prove this in the theory of Faddeev and Slavnov, again the
introduction of a preregulator is needed \cite{warr}.\\

\noindent
{\em Comment 2} : The theory is not anymore regularized for the higher loop
diagrams that contain a one loop subdiagram with external
PV~fields\cite{warr}.  I will not go further into this here.\\

\noindent
{\em Comment 3} : The theory is regularized at one loop level and is not
manifestly BRS invariant, because the mass term ---and only the mass term---
for the PV~fields breaks BRS invariance.  This does not mean however that
renormalizability is destroyed, because one can introduce at any higher
loop level a counterterm to restore gauge invariance.  Part of the one loop
counterterm for this regularization scheme without HCD is obtained in
\cite{JacBRST} for the Feynman gauge $\alpha=1$.  There, only the finite
part of the one loop counterterm was calculated without taking into account
the antifield dependent part.  Here I present the total one loop
counterterm and remark that there is no antifield dependence.  In
\cite{JacBRST} the heat kernel expansion is used to calculate the breaking
of BRS invariance ${\cal A} = (\Gamma ,\Gamma )$, where $\Gamma$ is the
Legendre transform w.r.t.  the sources of (\ref{Zreg}).  With the same
method one can easily obtain the infinite part.  ${\cal A}$ is then given
by
\begin{equation}
{\cal A} = \frac{1}{(4 \pi )^2} str(J a_2) -
\sum_{i=1}^{N} c_i M_i^2 \log \frac{M_i^2}{\mu ^2} str(J a_1) +
\sum_{i=1}^{N} c_i M_i^4 \log \frac{M_i^2}{\mu ^2} str(J a_0),
\label{anom}
\end{equation}
with $a_n$ the Seeley-DeWitt coefficients (cfr. \cite{JacBRST}).

For pure Yang-Mills (\ref{anom}) yields
\begin{eqnarray}
{\cal A}&=&- \frac{1}{(4 \pi)^2} \sum_{i=1}^{N} c_i M_i^2 \log
\frac{M_i^2}{\mu^2} str(c \partial_\alpha A^\alpha ) + \nonumber \\
&&\frac{1}{(4 \pi )^2 }\frac{1}{12} str[c(-4\partial^\nu A_\mu A_\nu A^\mu
\cdot + 4 \partial ^\nu A^\mu \cdot (\partial _\mu A_\nu - \partial_\nu
A_\mu) \nonumber \\
&&+ 4 (\partial^\nu A_\nu )^2 + 8 A^\mu \partial^\nu \partial _\mu A_\nu -4
A^\mu \Box A_\mu + 2 \Box \partial^\nu A_\nu )].
\end{eqnarray}

The one loop BRS invariance can be restored by a counterterm $\tilde M_1$
in the action such that ${\cal A} = (\tilde M_1,S)$.  The counterterm is
then
\begin{eqnarray}
\tilde M_1&=&\frac{1}{(4 \pi)^2}\left [ \frac{1}{2} \sum_{i=1}^{N} c_i
M_i^2 \log \frac{M_i^2}{\mu ^2} \, tr(A_\alpha A^\alpha) + \frac{1}{12}
tr({\textstyle\frac{3}{2}}(\partial_\mu A^\mu)^2 - {\textstyle
\frac{1}{2}}\partial_\mu A_\nu \cdot \partial^\mu A^\nu \right.\nonumber \\
&&\left.- 2 A^\nu A^\mu \partial _\mu A_\nu + {\textstyle \frac{3}{2}}A_\mu
A_\nu A^\mu A^\nu - {\textstyle \frac{1}{2}} A^2 A^2 ) \right].
\end{eqnarray}
If one now adds this counterterm to the action (\ref{Sgf}), $S = S_{gf} +
{\hbar} \tilde M_1$ one obtains the effective action by performing the
usual Legendre transformation.  This action we denote by
\begin{equation}
{\tilde \Gamma}^{(1)}(z^{cl}_0;g_0)=S(z^{cl}_0;g_0) + {\hbar} \tilde M_1,
\end{equation}
where $z^{cl} \equiv \{\Phi^{A\,cl},\Phi^*_A\}$.  The subscript $0$ denotes
that we have taken bare quantities.  This effective action is one-loop
BRS-invariant and satisfies a Zinn-Justin equation $({\tilde \Gamma}^{(1)},
{\tilde \Gamma}^{(1)})=0 + {\cal O}({\hbar}^2)$, which generates the 1-loop
Ward identities.

So one can perfectly adopt PV~regularization for gauge theories if one is
willing to destroy manifest gauge invariance and introduce counterterms.
So after introducing a counterterm, the theory is again manifestly gauge
invariant at one loop level.  Renormalization can be done in the usual way
and consistent results for the one loop $\beta$-function and anomalous
dimension of the fields are obtained.  This will be shown in section
\ref{oneloop}.\\

\noindent
{\em Comment 4} : It seems that there is no consistent gauge invariant
PV-regularization scheme.  In \cite{slav} a gauge invariant PV~action was
considered, which differs from \ref{Spv}.  However, then the the
regularization of the theory is destroyed and leads to inconsistent results
\cite{Ruiz}.

\end{subsection}

\begin{subsection}{Multiplicative renormalization and BV \label{ss:multbv}}
Suppose that one performs a multiplicative renormalization, as I will do in
section \ref{oneloop}.  Then one starts from the extended gauge fixed
action (\ref{Sgf}), where all quantities are taken to be `bare' quantities.
The gauge-fixing parameter $\alpha $ and the coupling constant $g$ get
renormalized by defining\footnote{Remark that we're not using the standard
notations for the $Z$-factors.  In our notation $Z_g=Z_1$ and $Z_A=Z_3$.}
\begin{eqnarray}
g_0&=& Z_g g_R,\\
\alpha_0&=& Z_\alpha \alpha_R.
\end{eqnarray}
For the fields and antifields we have
\begin{eqnarray}
A^\mu_0=Z^{1/2}_A A^\mu_R&\ ; \quad & A^*_{\mu \, 0}=\frac{1}{Z^{1/2}_A}
A^*_{\mu\, R}; \\
c_0=Z^{1/2}_c c_R &\ ; \quad & c^*_0=\frac{1}{Z^{1/2}_c} c^*_R; \\
b_0=Z^{1/2}_b b_R &\ ; \quad & b^*_0=\frac{1}{Z^{1/2}_b} b^*_R.
\end{eqnarray}

In fact this amounts to a canonical transformation of fields and antifields
with generating function $F(\Phi^A,\Phi'^*_A)=Z_\Phi \Phi^A \Phi'^*_A$ and
a redefinition of the parameters \cite{anselmi}.  Now certain relations
between the $Z$-factors can be derived.  In fact in (\ref{Sgf}) some terms
do not get radiative corrections, i.e.
$-\frac{1}{\alpha}b^*(1-\frac{\partial^2}{\lambda^2})\partial_\mu A^{\mu}$
and $-\frac{1}{2\alpha} {b^*}^2$.  From this one can conclude that
\begin{eqnarray}
Z_b=\frac{1}{Z_\alpha},\\
Z_b=\frac{1}{Z_A}.
\end{eqnarray}
This implies that one cannot choose $Z_b=Z_c$ as is done in previous
literature.  Because then one would have \begin{equation}
Z_c=Z_b=\frac{1}{Z_A} \end{equation} and this does not hold, as can be seen
from explicit one loop calculations in the following section.

\end{subsection}

\end{section}

\begin{section}{The one loop effective action \label{oneloop}}

In this section I will calculate the one loop effective action using two
regularization schemes.  First I use Pauli-Villars without introducing
Higher Covariant Derivatives.  In this scheme I calculate the $Z$-factors
and check the statements in section \ref{ss:multbv} by explicit
calculations.  With these $Z$-factors the renormalization group
coefficients are obtained, consistent with other regularization schemes.
Secondly I will present the calculations with the Higher Covariant
Derivative Pauli Villars scheme and obtain the same results.

\begin{subsection}{One loop effective action using Pauli-Villars
regularization \label{ss:pveffac}}

For the explicit calculations in this section I work in the Feynman gauge
$\alpha = 1$.  The Feynman rules are given in appendix A.

The divergent part (for $M_i^2 \rightarrow \infty$) of ${\tilde
\Gamma}^{(1)}$, which I denote by $\Gamma^{(1)}_{div}$ can be absorbed in
the $Z$-factors, from which the renormalization group coefficients are
computed.

We start by computing three 1PI Green functions, i.e.  the vacuum
polarization tensor $\Pi^{ab}_{\mu\nu}(p,M_i)= \langle A^a_\mu(-p)
A^b_\nu(p)\rangle $, the ghost-selfenergy $\Omega^{ab}(p,M_i)=\langle
b^a(-p) c^b(p)\rangle $ and the antighost-gluon-ghost vertex
$V_\mu^{abc}(p_1,p_2)=\langle b^a(p_2) A_\mu^b(p_1) c^c(-p_1-p_2)\rangle $.
The one loop corrections to $\Pi^{ab}_{\mu\nu}(p,M_i)$ are given by the
Feynman diagrams of fig.1 and result in
\begin{equation} \Pi^{ab\ (1)}_{\mu\nu}(p,M_i)=\lim_{M_i \rightarrow
\infty} \frac{g^2_0 c_v}{16 \pi^2} \left\{\frac{5}{3} \sum_{i=1}^N c_i
\log\left(\frac{M_i^2}{\kappa^2}\right) + \Pi^{(1)}_{fin} \right\}(p^2
g_{\mu\nu} - p_\mu p_\nu) \delta^{ab}
\label{vac}
\end{equation}
where $\kappa$ is the renormalization scale and $\Pi^{(1)}_{fin}$ collect
all finite terms for $M_i \rightarrow \infty$.  An explicit calculation is
lined out in appendix B.  For the other two 1PI Green functions
$\Omega^{ab}(p)$ and $V_\mu ^{abc}(p)$ the Feynman diagrams that give the
one loop corrections, are depicted in figs.  2 and 3 respectively.  They
result in
\begin{eqnarray}
\Omega^{ab\ (1)}(p)&=&\lim_{M_i \rightarrow \infty} \frac{g^2_0 c_v}{16
\pi^2} \left\{\frac{1}{2} \sum_{i=1}^N c_i
\log\left(\frac{M_i^2}{\kappa^2}\right) + \Omega^{(1)}_{fin} \right\} p^2
\delta^{ab}\label{gg}\\
V_\mu^{abc\ (1)}(p_1,p_2)&=&\lim_{M_i \rightarrow \infty}\frac{g^2_0
c_v}{16 \pi^2}\left\{ \frac{1}{2} \sum_{i=1}^N c_i
\log\left(\frac{M_i^2}{\kappa^2}\right) + V^{(1)}_{fin}\right\} g_0
f^a_{bc} i p_{2\,\mu}.\label{ggg}
\end{eqnarray}
In (\ref{vac}-\ref{ggg}) the $c_i$ satisfy the relations $\sum_{i=1}^N c_i=
-1$ and $\sum_{i=1}^N c_i M_i = 0$ (see appendix B).  From this functions
one can read of all the independent $Z$-factors:
\begin{eqnarray}
Z_A&=&1-\frac{5}{3}\, \lim_{M_i \rightarrow \infty}\frac{g_0^2 c_v}{16
\pi^2} \sum_{i=1}^N c_i \log\left (\frac{M_i^2}{\kappa^2}\right) \nonumber
\\ Z_b^{1/2} Z_c^{1/2}&=&1-\frac{1}{2}\, \lim_{M_i \rightarrow
\infty}\frac{g_0^2 c_v}{16 \pi^2} \sum_{i=1}^N c_i \log\left
(\frac{M_i^2}{\kappa^2}\right) \nonumber \\ Z_g Z_b^{1/2} Z_c^{1/2}
Z_A^{1/2}&=&1+\frac{1}{2}\, \lim_{M_i \rightarrow \infty}\frac{g_0^2
c_v}{16 \pi^2} \sum_{i=1}^N c_i \log\left (\frac{M_i^2}{\kappa^2}\right).
\end{eqnarray}
This yields \begin{eqnarray} Z_A^{1/2}\ =\ Z_\alpha&=&1-\frac{5}{6}\,
\lim_{M_i \rightarrow \infty} \frac{g_0^2 c_v}{16 \pi^2} \sum_{i=1}^N c_i
\log\left (\frac{M_i^2}{\kappa^2}\right)\nonumber \\
Z_b^{1/2}&=&1+\frac{5}{6}\, \lim_{M_i \rightarrow \infty}\frac{g_0^2
c_v}{16 \pi^2} \sum_{i=1}^N c_i \log\left (\frac{M_i^2}{\kappa^2}\right)
\nonumber \\
Z_c^{1/2}&=&1-\frac{4}{3}\, \lim_{M_i \rightarrow \infty}\frac{g_0^2
c_v}{16 \pi^2} \sum_{i=1}^N c_i \log\left (\frac{M_i^2}{\kappa^2}\right)
\nonumber\\ Z_g&=&1+\frac{11}{6}\, \lim_{M_i \rightarrow \infty}\frac{g_0^2
c_v}{16 \pi^2} \sum_{i=1}^N c_i \log\left
(\frac{M_i^2}{\kappa^2}\right).\label{zfac}
\end{eqnarray}
As a check on these results, I calculate the one loop correction to the
1PI-function $\langle A^{*a}_\mu(p) c^b(-p) \rangle$.  There is only the
contribution of the diagram in fig.  4.  The result is
\begin{equation}
\langle A^{*a}_\mu(p) c^b(-p) \rangle^{(1)} = - \lim_{M_i \rightarrow
\infty} \frac{g_0^2 c_v}{16 \pi^2}\left\{\frac{1}{2}\sum_{i=1}^N c_i
\log\left(\frac{M_i^2}{\kappa^2}\right) + {\rm fin} \right\} i p_\mu
\delta^{ab}
\end{equation}
and from this I derive that
\begin{equation}
\frac{Z^{1/2}_c}{Z^{1/2}_A} = 1 - \frac{1}{2}\, \lim_{M_i \rightarrow
\infty} \frac{g_0^2 c_v}{16 \pi^2} \sum_{i=1}^N c_i \log\left
(\frac{M_i^2}{\kappa^2}\right),
\end{equation}
which is in agreement with (\ref{zfac}).

Finally I come to the computation of the renormalization group
coefficients.  It is known that if one denotes by $\Gamma_R(p,\kappa,g)$
the renormalized 1PI Green functions for the Yang-Mills theory with $n_A$
external lines of the fields $\Phi^A$ in the Feynman gauge $\alpha=1$, the
renormalization group equation takes the form
\begin{equation}
\left[ \kappa \frac{\partial}{\partial \kappa} + \beta(g)
\frac{\partial}{\partial g} + \frac{1}{2} \sum_{A} \gamma_{\Phi^A}(g)
n_A\right] \Gamma_R(p,\kappa,g)=0.
\end{equation}
In multiplicative renormalization one then has that
\begin{eqnarray}
\beta(g)&\equiv&\kappa \frac{\partial g}{\partial \kappa} = \kappa g_0
\frac{\partial Z_g^{-1}}{\partial \kappa}\\ \gamma(g)&\equiv&-\kappa
\frac{\partial \ln(Z_\Phi)}{\partial \kappa}.
\end{eqnarray}
For the case I treat here, I obtain
\begin{eqnarray}
\beta(g)&=& -\frac{11}{3}\, \frac{c_v}{16 \pi^2} g^3 + {\cal
O}(g^4);\nonumber\\
\gamma_A(g)&=&\frac{10}{3}\, \frac{c_v}{16 \pi^2} g^2 + {\cal
O}(g^3);\nonumber\\
\gamma_b(g)&=&-\frac{10}{3}\, \frac{c_v}{16 \pi^2} g^2
+ {\cal O}(g^3);\nonumber\\
\gamma_c(g)&=&\frac{16}{3}\, \frac{c_v}{16
\pi^2} g^2 + {\cal O}(g^3).\label{resrg}
\end{eqnarray}
These are the standard results for $\beta(g)$ and $\gamma_A(g)$.
\end{subsection}

\begin{subsection}{One loop effective action using HCD-PV regularization
\label{ss:effacHCD}}
Now I use the full regularization scheme defined in section \ref{s:regac}
to compute the one-loop 1PI-functions.\\
The strategy is as follows.  I have redone the calculations of the previous
subsection without knowing the counterterm to restore BRS-invariance.  This
counterterm has to be local.  The contributions to the $Z$-factor are
non-local.  So if one is only interested in these coefficients, one doesn't
need the exact form of the counterterm.  In order to do the very extensive
algebra, I used the Mathematica package HIP \cite{HIP}.  I took the limits
in the order $\lim_{\Lambda\rightarrow\infty} \lim_{{M_i}\rightarrow\infty}
$.  The results are then
\begin{eqnarray} \Pi^{ab\
(1)}_{\mu\nu}(p,M_i)&=&\lim_{\Lambda\rightarrow\infty}
\lim_{{M_i}\rightarrow\infty} \frac{g_0^2 c_v}{16 \pi^2}
\left\{-\frac{19}{12} \log\left(\frac{\Lambda^2}{\kappa^2}\right)
+\frac{1}{12}\sum_{i=1}^N c_i
\log\left(\frac{M_i^2}{\kappa^2}\right)\right\}p^2
g_{\mu\nu}\delta^{ab}\nonumber\\
&&+\lim_{\Lambda\rightarrow\infty} \lim_{{M_i}\rightarrow\infty}
\frac{g_0^2 c_v}{16 \pi^2} \left\{ \frac{11}{6}
\log\left(\frac{\Lambda^2}{\kappa^2}\right) +\frac{1}{6}\sum_{i=1}^N c_i
\log\left(\frac{M_i^2}{\kappa^2}\right)\right\}p_\mu p_\nu \delta^{ab}
\nonumber\\
&&+ \kappa{\rm-independent} + {\rm fin}, \label{vac2}\\
\Omega^{ab\ (1)}(p)&=&-\lim_{\Lambda \rightarrow \infty} \frac{g^2_0
c_v}{16 \pi^2}
\left\{\frac{1}{2}\log\left(\frac{\Lambda^2}{\kappa^2}\right)\right\} p^2
\delta^{ab} + {\rm fin}, \label{gg2}\\ V_\mu^{abc\
(1)}(p,k)&=&-\lim_{\Lambda \rightarrow \infty}\frac{g^2_0 c_v}{16
\pi^2}\left\{ \frac{1}{2} \log\left(\frac{\Lambda}{\kappa^2}\right)\right\}
g_0 f^a_{bc} i p_\mu+ {\rm fin}.\label{ggg2}
\end{eqnarray}
One can easily see that equations (\ref{vac2}-\ref{ggg2}) lead to the same
Renormalization Group coefficients (\ref{resrg}).  So the HCD-PV
regularization scheme as it is presented here leads to consistent results.
\end{subsection}
\end{section}

\begin{section}{Conclusions \label{s:conclusions}}
In this paper I have derived a Higher Covariant Derivative Pauli Villars
regularization scheme for pure Yang-Mills within the context of the Batalin
Vilkovisky formalism.  This was done in two steps.  First I introduced the
Higher Covariant derivatives and did the gauge fixing.  This regularizes
all higher loop diagrams.  In order to regularize one loop (sub)diagrams
PV~fields are added to the theory.  Here BRS invariance is broken
explicitly, but the counterterm for the case of PV has been presented.
When calculating the one loop effective action everything is consistent
with other regularization schemes.  The BV scheme implies a preferred
choice for the $Z$-factors.

The remaining question, which I didn't address here, is ``What about the
higher loop diagrams?" It is clear that one has to introduce more PV~fields
for the higher loop diagrams, even an infinite number of generations.
There is a priori no objection to that because PV is applied in a
perturbative context.  One drawback is that when you introduce second order
PV~fields they not only have to couple to first order PV~fields but also to
the original fields.  The use of Higher Covariant Derivatives makes the
algebra very complicated (cfr.  appendix A) and doesn't regularize all
higher loop diagrams.  Therefore, in spite of the consistency exhibited in
this paper, one could have some doubt about the usefulness of Higher
Covariant Derivative regularization.

\end{section}

\vspace{1cm}
\noindent
{\large \bf Acknowledgements}

I wish to thank W.  Troost and A.  Van Proeyen for introducing me to this
problem and for useful suggestions.
\vspace{1cm}
\appendix

\begin{section}{Feynman rules}
The Feynman rules associated with (\ref{Zreg}) are:
\begin{center}
{\large \bf Propagators}\\

\begin{picture}(330,40)(0,0)
\Photon(10,20)(70,20){3}{6}\Vertex(10,20){2}\Vertex(70,20){2}
\Text(10,12)[tc]{$A^a_\mu$} \Text(40,30)[bc]{${\ds p\atop \rightarrow}$}
\Text(70,12)[tc]{$A^b_\nu$}
\Text(200,20)[l]{$ \ds  \frac{\delta^{ab}\Lambda^4
g_{\mu\nu}}{p^2\,(p^2+\Lambda^2)^2}$}
\end{picture}\end{center}
\begin{center}\begin{picture}(330,60)(0,0)
\Gluon(10,20)(70,20){4}{6}\Vertex(10,20){2}\Vertex(70,20){2}
\Text(10,12)[tc]{$A^{~\,a}_{i\,\mu}$}
\Text(40,30)[bc]{$\ds p\atop \rightarrow $}
\Text(70,12)[tc]{$A^{~\,b}_{i\,\nu}$}
\Text(200,20)[l]{$ \ds  \frac{\delta^{ab}\Lambda^4
g_{\mu\nu}}{(p^2+M_i^2)\,(p^2+\Lambda^2)^2}$}
\end{picture}\end{center}
\begin{center}\begin{picture}(330,60)(0,0)
\DashArrowLine(10,20)(70,20){6}
\Vertex(10,20){2}\Vertex(70,20){2}
\Text(10,12)[tc]{$b^a$} \Text(40,30)[bc]{$\ds p$}
\Text(70,12)[tc]{$c^b$}
\Text(200,20)[l]{$\ds  \frac{\delta^{ab} \Lambda^2}{p^2\,(p^2 +
\Lambda^2)}$}
\end{picture}\end{center}
\begin{center}\begin{picture}(330,60)(0,0)
\ArrowLine(10,20)(70,20)\Vertex(10,20){2}\Vertex(70,20){2}
\Text(10,12)[tc]{$b^a_i$} \Text(40,30)[bc]{$p$}
\Text(70,12)[tc]{$c^b_i$}
\Text(200,20)[l]{$\ds  \frac{\delta^{ab} \Lambda^2}{(p^2 + M_i^2)\,(p^2
+ \Lambda^2)}$}
\end{picture}\end{center}
These Feynman rules are given for the Feynman gauge $\alpha=1$.  As we
pointed out in section \ref{s:regac}, the denominators factorize in factors
that are quadratic in $p$.  This gives the opportunity to use standard
manipulations (feynman-parameters etc.) to calculate diagrams (cfr.
Appendix B).\\
\vskip1cm
\begin{center}{\large \bf Vertices}\end{center}

\begin{center}\begin{picture}(410,105)(0,0)
\Photon(0,22.5)(30,40){2}{4} \LongArrow(4.5,33)(14,39)
\Photon(30,40)(60,22.5){2}{4} \LongArrow(46.5,22)(37,28)
\Photon(30,40)(30,75){2}{4} \LongArrow(36,64)(36,53)
\Vertex(30,40){2}
\Text(30,86)[cb]{$A^{a_1}_{\mu_1}$} \Text(39,59)[cl]{$p_1$}
\Text(3,14)[tr]{$A^{a_2}_{\mu_2}$} \Text(8,40)[cr]{$p_2$}
\Text(55,14)[tl]{$A^{a_3}_{\mu_3}$} \Text(45,20)[tr]{$p_3$}
\Gluon(110,22.5)(140,40){3}{4}\Gluon(140,40)(170,22.5){3}{4}
\Photon(140,40)(140,75){2}{4}\Vertex(140,40){2}
\Text(140,86)[cb]{$A^{a_1}_{\mu_1}$}
\Text(120,14)[tr]{$A_j{}^{a_2}_{\mu_2}$}
\Text(160,14)[tl]{$A_j{}^{a_3}_{\mu_3}$}
\Text(149,59)[cl]{$p_1$} \LongArrow(114.5,33)(124,39)
\Text(118,40)[cr]{$p_2$} \LongArrow(156.5,22)(147,28)
\Text(155,20)[tr]{$p_3$} \LongArrow(146,64)(146,53)
\Text(210,70)[l]{$ \ds - {ig\over\Lambda^4} {\rm \bf S}_3 \Big\{
       f^{a_1}_{a_2a_3}\, \Big[ \,\Lambda^4 \, p_{1\mu_2}\,
g_{\mu_3\mu_1} - p_1^4 p_{1\mu_2}g_{\mu_1\mu_2}$}
\Text(240,40)[l]{$ \ds  + \Lambda^2 \, p_{1\mu_2}\, (p_1^2 \,
g_{\mu_3\mu_1} + p_{1\mu_3} p_{3\mu_1}) $}
\Text(240,10)[l]{$ \ds   \, p_1^2\, (p_3 - p_1)_{\mu_2} \, \big(\,
       p_{1\mu_3} \, p_{3\mu_1}\! - p_1 \cdot p_3\, g_{\mu_1\mu_3}
           \big) \,\Big] \Big\} $}
\end{picture}\end{center}
where ${\rm \bf S}_3$ is the symmetrization operator with respect to the
indices 1, 2 and 3.

\begin{center}\begin{picture}(410,220)(0,0)
\DashArrowLine(0,112.5)(30,130){6}\DashArrowLine(30,130)(60,112.5){6}
\Photon(30,130)(30,165){2}{4}\Vertex(30,130){2}
\LongArrow(46.5,112)(37,118)
\Text(30,176)[cb]{$A^a_\mu$} \Text(3,104)[tr]{$b^b$}
\Text(55,104)[tl]{$c^c$} \Text(45,110)[tr]{$p$}
\ArrowLine(110,112.5)(140,130)\ArrowLine(140,130)(170,112.5)
\Photon(140,130)(140,165){2}{4}\Vertex(140,130){2}
\LongArrow(156.5,112)(147,118)
\Text(140,176)[cb]{$A^a_\mu$} \Text(113,104)[tr]{$b^b_i$}
\Text(165,104)[tl]{$c^c_i$} \Text(155,110)[tr]{$p$}
\DashArrowLine(0,22.5)(30,40){6}\ArrowLine(30,40)(60,22.5)
\Gluon(30,40)(30,75){2}{4}\Vertex(30,40){2}
\LongArrow(46.5,22)(37,28)
\Text(30,86)[cb]{$A^a_{i\,\mu}$} \Text(3,14)[tr]{$b^b$}
\Text(55,14)[tl]{$c^c_i$} \Text(45,20)[tr]{$p$}
\ArrowLine(110,22.5)(140,40)\DashArrowLine(140,40)(170,22.5){6}
\Gluon(140,40)(140,75){2}{4}\Vertex(140,40){2}
\LongArrow(156.5,22)(147,28)
\Text(140,86)[cb]{$A^a_{i\,\mu}$} \Text(113,14)[tr]{$b^b_i$}
\Text(165,14)[tl]{$c^c$} \Text(155,20)[tr]{$p$}
\Text(210,90)[l]{$\ds -\,
\frac{ig}{\Lambda^2}f^a_{bc}\,(\Lambda^2 + p^2)\,p_\mu$}
\end{picture}\end{center}
\begin{center}\begin{picture}(170,90)(0,0)
\Photon(0,0)(25,25){2}{4}
\LongArrow(10,3)(18,11) \Text(20,2)[c]{$p_2$}
\Photon(25,25)(0,50){2}{4}
\LongArrow(3,40)(11,32) \Text(2,30)[c]{$p_1$}
\Photon(25,25)(50,0){2}{4}
\LongArrow(47,10)(39,18) \Text(52,18)[c]{$p_3$}
\Photon(25,25)(50,50){2}{4}
\LongArrow(40,47)(32,39) \Text(30,48)[c]{$p_4$}
\Vertex(25,25){2}
\Text(3,60)[br]{$A^{a_1}_{\mu_1}$}\Text(3,-4)[tr]{$A^{a_2}_{\mu_2}$}
\Text(45,-4)[tl]{$A^{a_3}_{\mu_3}$}\Text(45,60)[bl]{$A^{a_4}_{\mu_4}$}
\Gluon(110,0)(135,25){3}{4}\Gluon(135,25)(160,0){3}{4}
\LongArrow(120,3)(128,11) \Text(130,2)[c]{$p_2$}
\LongArrow(113,40)(121,32) \Text(112,30)[c]{$p_1$}
\LongArrow(157,10)(149,18) \Text(162,18)[c]{$p_3$}
\LongArrow(150,47)(142,39) \Text(140,48)[c]{$p_4$}
\Photon(135,25)(110,50){2}{4}\Photon(135,25)(160,50){2}{4}
\Vertex(135,25){2}
\Text(117,60)[br]{$A^{a_1}_{\mu_1}$}
\Text(120,-4)[tr]{$A^{a_2}_{i\,\mu_2}$}
\Text(150,-4)[tl]{$A^{a_3}_{i\,\mu_3}$}
\Text(155,60)[bl]{$A^{a_4}_{\mu_4}$}
\end{picture}\end{center}

\begin{eqnarray}
-\frac{g^2}{4 \Lambda^4}{\rm \bf S}_4 &\Big\{& f^{a_1}_{e a_2} f^e_{a_3
a_4} \Big[g_{\mu_1\mu_3} g_{\mu_2\mu_4} \Big(\Lambda^4 - 4 \Lambda^2(2p_1 +
p_2)\cdot p_3 + (p_1 + p_2)^2 (p_3 + p_4)^2 \Big)\nonumber\\[8pt]
& &-2 \Lambda^2 \Big( g_{\mu_2\mu_3}p_{1\mu_4}p_{3\mu3} +
g_{\mu_1\mu_4}p_{1\mu_3}(2p_1 + p_2)_{\mu_2}\Big)\nonumber\\[8pt]
& &-4 (2p_1 + p_2)_{\mu_2} g_{\mu_1\mu_4} p_{1\mu_3}\Big( (p_1+p_2)^2 +
p_1^2\Big)\nonumber\\[8pt]
& &+ 2 p_1\cdot p_4 g_{\mu_1\mu_3} \Big( 2 p_1^2 g_{\mu_2\mu_3} -
p_{2\mu_2} p_{3\mu_3} + 4 p_{1\mu_2} (p_1 +
p_2)_{\mu_3}\Big)\nonumber\\[8pt]
& &-4 p_1^2 g_{\mu_2\mu_3} p_{1\mu_4} p_{4\mu_1} + 2 p_{1\mu_4} p_{4\mu_1}
(p_{2\mu_2}p_{3\mu_3} - 4 p_{2\mu_3} p_{1\mu_2} - 4 p_{1\mu_2} p_{1\mu_3})
\Big] \Big\}.\nonumber
\end{eqnarray}
Here also ${\rm \bf S}_4$ means symmetrization with respect to 1,2,3 and 4.

I do not exhibit the higher order vertices here, because they don't show
up in the calculations at one-loop order.
\end{section}

\begin{section}{A sample computation}
In this appendix I will analyse the explicit calculation of the one loop
Feynman diagrams.  My mean point here is to show that one can stay at
space-time dimension four during the whole calculation.  As an example I
take the first line of fig.  1.

After doing the algebra the momentum integral is
\begin{eqnarray}
I&=&\frac{1}{2} f^a_{cd} f^b_{cd} \int \frac{d^4 k}{(2 \pi)^4}G(\mu,\nu,p,k) \nonumber\\
& &\quad \sum_{i=1}^N c_i \Big(\frac{1}{(k^2 + M_i^2)(k^2 +
\Lambda^2)^2((k+p)^2 + M_i^2)((k+p)^2 + \Lambda^2)^2} -\nonumber\\
& &\quad\quad  \frac{1}{k^2 (k^2 +
\Lambda^2)^2(k+p)^2((k+p)^2 + \Lambda^2)^2}\Big).
\end{eqnarray}
In order to perform this step, we have to demand that
\begin{equation}
\sum_{i=1}^N c_i=-1.
\end{equation}
Because the denominators are factorizing one can now use the standard
Feynman parameter procedure to combine the factors,
\begin{eqnarray}
I&=&\frac{6}{2} f^a_{cd} f^b_{cd} \int \frac{d^4 k}{(2
\pi)^4} \int_0^1 dx \int_0^1 dy G(\mu,\nu,p,k) y(1-y)\nonumber\\
& &\quad \sum_{i=1}^N c_i \Big(\frac{1}{(k^2 + 2x k\cdot p + x
p^2 + M_i^2)^2 (k^2 + 2y k\cdot p + y p^2
+ \Lambda^2)^4} -\nonumber\\
& &\quad\quad  \frac{1}{(k^2 + 2 x k\cdot p + x p^2)^2 (k^2 + 2y k\cdot p +
y p^2 \Lambda^2)^4}\Big).
\end{eqnarray}
In order to increase the degree of the denominator we combine the terms
through,
\begin{equation}
\frac{1}{(k+a)^n} - \frac{1}{(k+b)^n} = n \int_a^b
\frac{dx}{(k+x)^{n+1}}.
\label{trick}
\end{equation}
Applying this twice, demanding that
\begin{equation}
\sum_{i=1}^N c_i M_i^2 = 0,
\end{equation}
and using another Feynman parameter,
you arrive at
\begin{eqnarray}
I&=&\frac{\Gamma(8)}{2} f^a_{cd} f^b_{cd} \int \frac{d^4 k}{(2
\pi)^4} \int_0^1 dx \int_0^1 dy \sum_{i=1}^N c_i \int_0^{M_i^2}da
\int_{\kappa^2}^a db \int_0^1 dz G(\mu,\nu,p,k)z^3(1-z)^3 y(1-y)\nonumber\\
& &\quad \Big(\frac{1}{(k^2 + 2 z x k\cdot p + z x p^2 + 2(1-z)y k\cdot p +
 (1-z) y p^2 + z b +(1-z \Lambda^2)^8}\Big).
\end{eqnarray}
Note that in the last time you apply (\ref{trick}) the renormalization
scale $\kappa$ comes in.  Finally one shifts $k$ to $k'=k - (zx (1-z)y)p$.
This can be done because the $k$-integral is now finite.  Now all integrals
can be interchanged and performed starting with the momentum integral.
Once again I'd like to stress that all manipulations were done in four
dimensions.
\end{section}

\newpage
\section*{Figures}
\begin{center}
\begin{picture}(460,180)(0,0)
\Photon(50,100)(90,100){3}{3}
\Photon(170,100)(210,100){3}{3}
\PhotonArc(130,100)(40,0,180){3}{8.5}
\PhotonArc(130,100)(40,180,360){3}{8.5}
\ArrowArc(130,100)(30,60,120)
\ArrowArc(130,100)(30,240,300)
\Text(130,125)[t]{$k$}
\Text(130,75)[b]{$k+p$}
\Vertex(90,100){2} \Vertex(170,100){2}
\Text(50,86)[l]{$A^a_{\mu}$} \Text(210,86)[r]{$A^b_\mu$}
\Text(70,105)[b]{$\ds p \atop \rightarrow$}
\Text(190,105)[b]{$\ds p \atop \rightarrow$}
\Text(0,100)[l]{$\ds \int d^4 k$}
\Photon(280,100)(320,100){3}{3}
\Photon(400,100)(440,100){3}{3}
\GlueArc(360,100)(40,0,180){4}{8}
\GlueArc(360,100)(40,180,360){4}{8}
\Vertex(320,100){2} \Vertex(400,100){2}
\ArrowArc(360,100)(30,60,120)
\ArrowArc(360,100)(30,240,300)
\Text(360,125)[t]{$k$}
\Text(360,75)[b]{$k+p$}
\Text(240,100)[l]{$\ds + \sum_{i=1}^N c_i$}
\Text(280,86)[l]{$A^a_\mu$} \Text(435,86)[r]{$A^b_\mu$}
\Text(300,105)[b]{$\ds p \atop \rightarrow$}
\Text(420,105)[b]{$\ds p \atop \rightarrow$}
\end{picture}
\begin{picture}(460,115)(0,0)
\Text(0,85)[l]{$\ds \int d^4 k$}
\Photon(55,40)(175,40){3}{8.5} \PhotonArc(115,85)(40,270,269){3}{15.5}
\Vertex(115,43.5){2}
\Text(50,45)[lb]{$A^a_\mu$} \Text(170,45)[lb]{$A^b_\mu$}
\ArrowArc(115,85)(30,60,120)
\Text(115,110)[t]{$k$}
\Text(70,35)[t]{$\ds \rightarrow$}
\Text(160,35)[t]{$\ds \rightarrow$}
\Text(70,30)[t]{$\ds p$}
\Text(160,30)[t]{$\ds p$}
\Photon(295,40)(425,40){3}{8.5} \GlueArc(360,85)(40,-90,270){4}{16}
\Vertex(360,43.5){2}
\Text(240,85)[l]{$\ds + \sum_{i=1}^N c_i$}
\Text(290,45)[lb]{$A^a_\mu$} \Text(420,45)[lb]{$A^b_\mu$}
\ArrowArc(360,85)(30,60,120)
\Text(360,110)[t]{$k$}
\Text(310,35)[t]{$\ds \rightarrow$}
\Text(410,35)[t]{$\ds \rightarrow$}
\Text(310,30)[t]{$\ds p$}
\Text(410,30)[t]{$\ds p$}
\end{picture}
\begin{picture}(460,140)(0,0)
\Photon(50,100)(90,100){3}{3}
\Photon(170,100)(210,100){3}{3}
\DashArrowArc(130,100)(40,0,180){6}
\DashArrowArc(130,100)(40,180,360){6}
\Text(130,130)[t]{$k$}
\Text(130,65)[b]{$k+p$}
\Vertex(90,100){2} \Vertex(170,100){2}
\Text(50,86)[l]{$A^a_{\mu}$} \Text(210,86)[r]{$A^b_\mu$}
\Text(70,105)[b]{$\ds p \atop \rightarrow$}
\Text(190,105)[b]{$\ds p \atop \rightarrow$}
\Text(0,100)[l]{$\ds \int d^4 k$}
\Photon(280,100)(320,100){3}{3}
\Photon(400,100)(440,100){3}{3}
\ArrowArc(360,100)(40,0,180)
\ArrowArc(360,100)(40,180,360)
\Vertex(320,100){2} \Vertex(400,100){2}
\Text(360,130)[t]{$k$}
\Text(360,65)[b]{$k+p$}
\Text(240,100)[l]{$\ds + \sum_{i=1}^N c_i$}
\Text(280,86)[l]{$A^a_\mu$} \Text(435,86)[r]{$A^b_\mu$}
\Text(300,105)[b]{$\ds p \atop \rightarrow$}
\Text(420,105)[b]{$\ds p \atop \rightarrow$}
\end{picture}
\begin{picture}(460,40)(0,0)
\Photon(50,20)(170,20){3}{9}
\BBoxc(110,20)(10,10)
\Text(50,6)[l]{$A^a_\mu$} \Text(170,6)[r]{$A^b_\nu$}
\Text(80,25)[b]{$\ds p \atop \rightarrow$}
\Text(140,25)[b]{$\ds p \atop \rightarrow$}
\Text(200,20)[l]{{\sl counterterm to restore BRS-invariance}}
\end{picture}\\[.5cm]
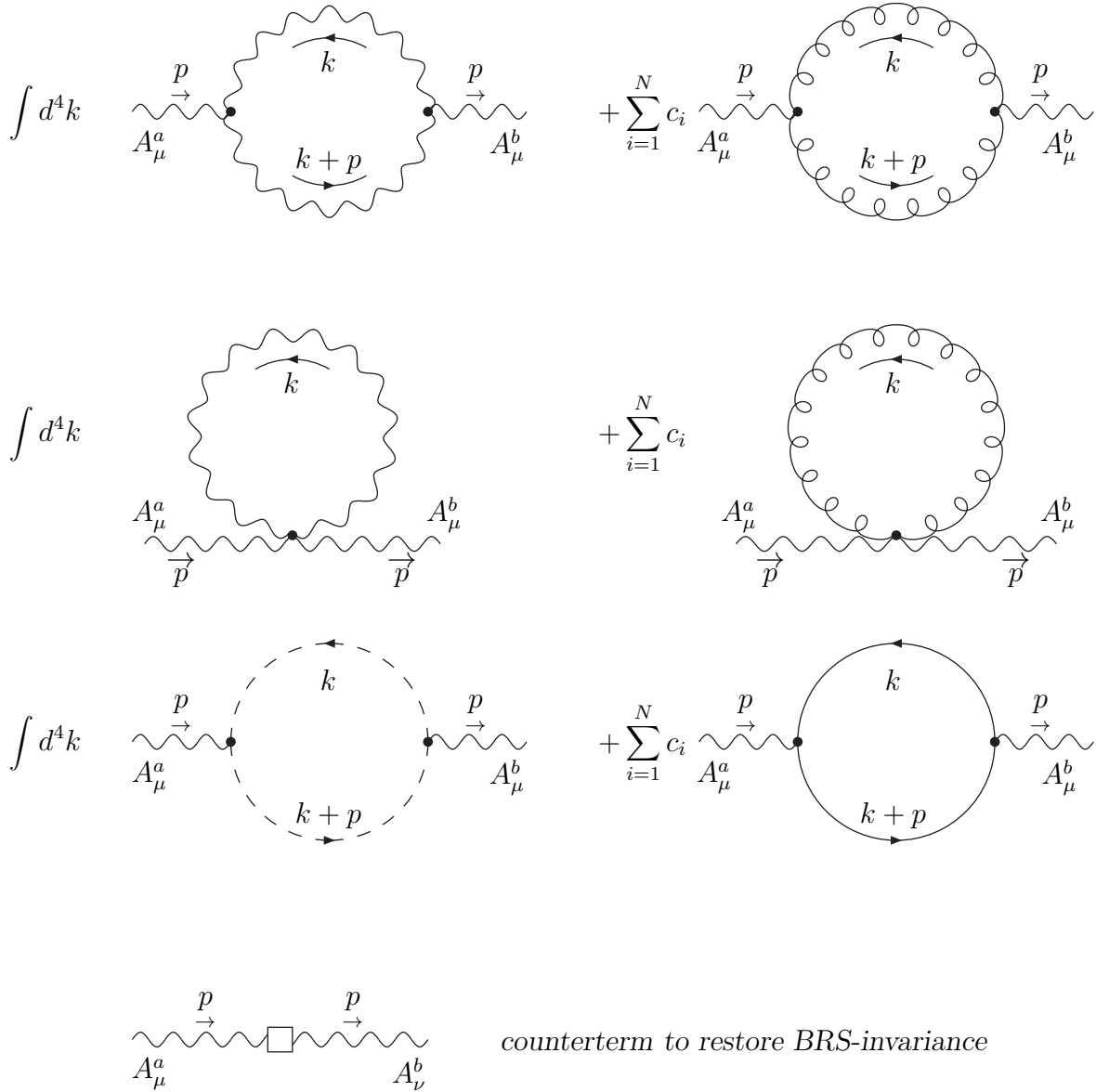
fig 1.: {\sl One loop contributions to the vacuum polarization tensor}

\vspace{1cm}

\begin{picture}(460,50)(0,0)
\Text(0,20)[l]{$\ds \int d^4 k$}
\DashArrowLine(50,20)(90,20){6} \DashArrowLine(170,20)(210,20){6}
\DashArrowLine(90,20)(170,20){6}\PhotonArc(130,20)(40,0,180){3}{8.5}
\Vertex(90,20){2} \Vertex(170,20){2}
\Text(80,10)[b]{$c^b$} \Text(180,10)[b]{$b^a$}
\Text(70,25)[b]{$p$} \Text(190,25)[b]{$p$}
\Text(130,15)[t]{$k+p$} \ArrowArc(130,20)(30,60,120) \Text(130,45)[t]{$k$}
\DashArrowLine(280,20)(320,20){6} \DashArrowLine(400,20)(440,20){6}
\ArrowLine(320,20)(400,20) \GlueArc(360,20)(40,0,180){3}{8}
\Vertex(320,20){2} \Vertex(400,20){2}
\Text(240,20)[l]{$\ds + \sum_{i=1}^N c_i$}
\Text(312,10)[b]{$c^b$} \Text(410,10)[b]{$b^a$}
\Text(300,25)[b]{$p$} \Text(420,25)[b]{$p$}
\Text(360,15)[t]{$k+p$} \ArrowArc(360,20)(30,60,120) \Text(360,45)[t]{$k$}
\end{picture}\\[5mm]
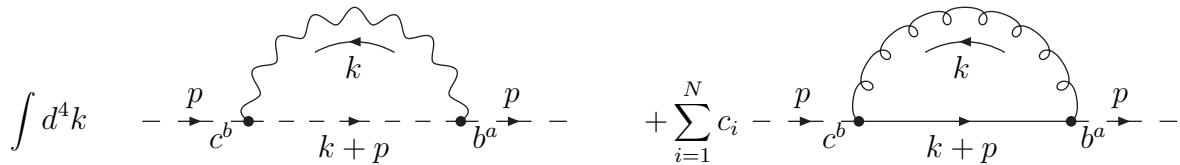
fig 2.: {\sl One loop contributions to the ghost selfenergy}

\begin{picture}(460,150)(0,0)
\Text(0,75)[l]{$\ds \int d^4 k$} \DashArrowLine(45,15)(79.6,35){6}
\DashArrowLine(79.6,35)(114.2,95){6} \DashArrowLine(114.2,95)(148.8,35){6}
\DashArrowLine(148.8,35)(183.4,15){6}
\Photon(79.6,35)(148.8,35){3}{5}\Photon(114.2,95)(114.2,135){3}{3}
\Vertex(79.6,35){2}\Vertex(148.8,35){2}\Vertex(114.2,95){2}
\Text(62.6,32)[r]{$c^b$} \Text(169.8,32)[l]{$b^c$}
\Text(119.2,135)[l]{$A_\mu^a$} \LongArrow(119,30)(109,30)
\Text(114,28)[t]{$k + p_2$} \LongArrow(93,70)(98.2,80) \Text(90,75)[r]{$k -
p_1$} \LongArrow(129.8,80)(135,70) \Text(138,75)[l]{$k$}
\LongArrow(120,125)(120,115) \Text(125,120)[l]{$p_1$}
\LongArrow(169,15)(161,19) \Text(165,10)[]{$p_2$} \Text(200,75)[l]{$\ds +
\sum_{i=1}^N c_i$} \DashArrowLine(250,15)(284.6,35){6}
\ArrowLine(284.6,35)(319.2,95) \ArrowLine(319.2,95)(353.8,35)
\DashArrowLine(353.8,35)(388.4,15){6} \Gluon(284.6,35)(353.8,35){3}{5}
\Photon(319.2,95)(319.2,135){3}{3}
\Vertex(284.6,35){2}\Vertex(353.8,35){2}\Vertex(319.2,95){2}
\Text(267.6,32)[r]{$c^b$} \Text(374.8,32)[l]{$b^b$}
\Text(324.2,135)[l]{$A_\mu^a$} \LongArrow(324,30)(314,30)
\Text(319,28)[t]{$k + p_2$} \LongArrow(298,70)(303.2,80)
\Text(295,75)[r]{$k - p_1$} \LongArrow(334.8,80)(340,70)
\Text(343,75)[l]{$k$} \LongArrow(325,125)(325,115) \Text(330,120)[l]{$p_1$}
\LongArrow(374,15)(366,19) \Text(370,10)[]{$p_2$}
\end{picture}\\

\begin{picture}(460,150)(0,0)
\Text(0,75)[l]{$\ds \int d^4 k$} \DashArrowLine(45,15)(79.6,35){6}
\Photon(79.6,35)(114.2,95){3}{5} \Photon(114.2,95)(148.8,35){3}{5}
\DashArrowLine(148.8,35)(183.4,15){6}
\DashArrowLine(79.6,35)(148.8,35){6}\Photon(114.2,95)(114.2,135){3}{3}
\Vertex(79.6,35){2}\Vertex(148.8,35){2}\Vertex(114.2,95){2}
\Text(62.6,32)[r]{$c^b$} \Text(169.8,32)[l]{$b^c$}
\Text(119.2,135)[l]{$A_\mu^a$} \LongArrow(119,30)(109,30)
\Text(114,28)[t]{$k + p_2$} \LongArrow(93,70)(98.2,80) \Text(90,75)[r]{$k -
p_1$} \LongArrow(129.8,80)(135,70) \Text(138,75)[l]{$k$}
\LongArrow(120,125)(120,115) \Text(125,120)[l]{$p_1$}
\LongArrow(169,15)(161,19) \Text(165,10)[]{$p_2$} \Text(200,75)[l]{$\ds +
\sum_{i=1}^N c_i$} \DashArrowLine(250,15)(284.6,35){6}
\Gluon(284.6,35)(319.2,95){3}{5} \Gluon(319.2,95)(353.8,35){3}{5}
\DashArrowLine(353.8,35)(388.4,15){6} \ArrowLine(284.6,35)(353.8,35)
\Photon(319.2,95)(319.2,135){3}{3}
\Vertex(284.6,35){2}\Vertex(353.8,35){2}\Vertex(319.2,95){2}
\Text(267.6,32)[r]{$c^b$} \Text(374.8,32)[l]{$b^b$}
\Text(324.2,135)[l]{$A_\mu^a$} \LongArrow(324,30)(314,30)
\Text(319,28)[t]{$k + p_2$} \LongArrow(298,70)(303.2,80)
\Text(295,75)[r]{$k - p_1$} \LongArrow(334.8,80)(340,70)
\Text(343,75)[l]{$k$} \LongArrow(325,125)(325,115) \Text(330,120)[l]{$p_1$}
\LongArrow(374,15)(366,19) \Text(370,10)[]{$p_2$}
\end{picture}\\[.5cm]
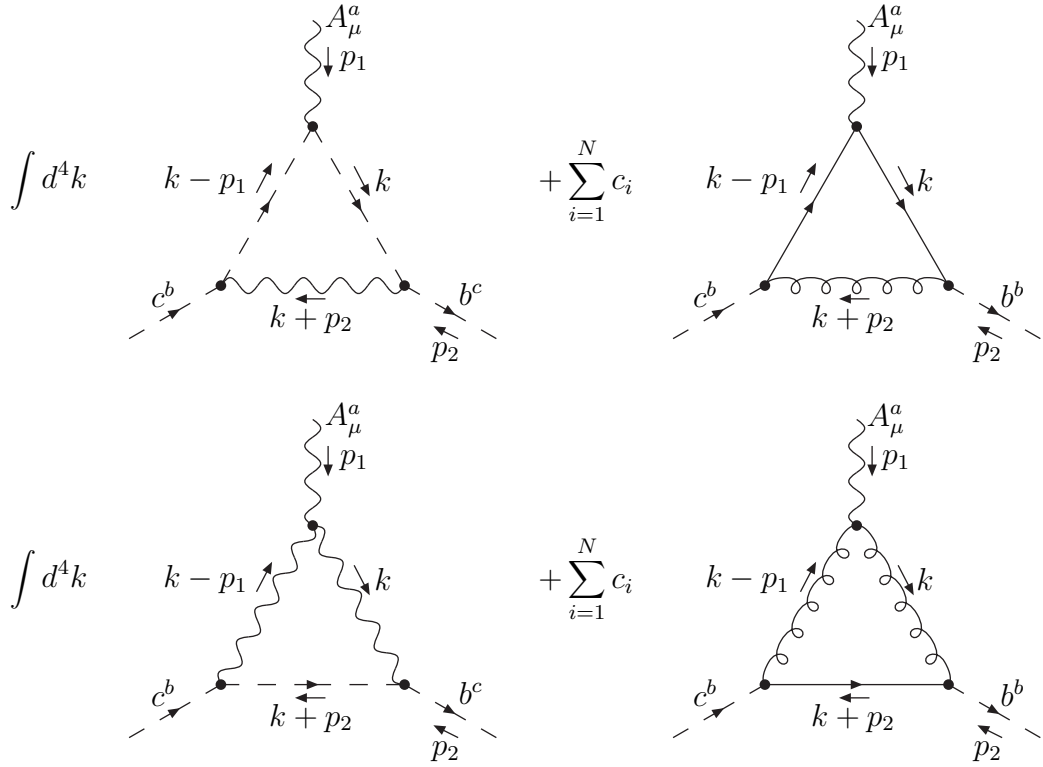
fig.  3: {\sl One loop contributions to the ghost-gluon-ghost vertex.}\\

\vspace{2cm}

\begin{picture}(460,50)(0,0)
\Text(0,20)[l]{$\ds \int d^4 k$} \Line(50,19)(90,19)
\Line(50,21)(90,21)\DashArrowLine(210,20)(170,20){6}
\DashArrowLine(170,20)(90,20){6}\PhotonArc(130,20)(40,0,180){3}{8.5}
\Vertex(90,20){2} \Vertex(170,20){2} \Text(50,10)[t]{$A^{*a}_{\mu}$}
\Text(180,10)[b]{$c^b$} \Text(70,25)[b]{$\ds p \atop \rightarrow$}
\Text(190,25)[b]{$\ds p \atop \leftarrow$} \Text(130,15)[t]{$k-p$}
\ArrowArc(130,20)(30,60,120) \Text(130,40)[t]{$k$} \Line(280,19)(320,19)
\Line(320,21)(280,21)\DashArrowLine(440,20)(400,20){6}
\ArrowLine(400,20)(320,20) \GlueArc(360,20)(40,0,180){3}{8}
\Vertex(320,20){2} \Vertex(400,20){2} \Text(230,20)[l]{$\ds + \sum_{i=1}^N
c_i$} \Text(280,10)[t]{$A^{*a}_{\mu}$} \Text(410,10)[b]{$c^b$}
\Text(300,25)[b]{$\ds p \atop \rightarrow$} \Text(420,25)[b]{$\ds p \atop
\leftarrow$} \Text(360,15)[t]{$k-p$} \ArrowArc(360,20)(30,60,120)
\Text(360,45)[t]{$k$}
\end{picture}\\[.5cm]
fig 4.: {\sl One loop contributions to} $\langle A^{*a}_{\mu} \partial^\mu
c^a \rangle$
\end{center}
\end{document}